\documentclass[prl,twocolumn,aps]{revtex4}
\usepackage{hyperref}
\usepackage{graphicx}
\usepackage{subfigure}
\usepackage{amsmath}
\usepackage[squaren]{SIunits}
\begin{document}
\title{Pattern formation upon femtosecond laser ablation of transparent dielectrics}
\author{Ionu\c{t} Georgescu}%
\altaffiliation[Current address ]{Max-Planck-Institut f\"{u}r Physik komplexer Systeme, N\"{o}thnitzer Str. 38, 01187 Dresden, Germany.}%
\email{george@pks.mpg.de}

\author{Michael Bestehorn}
\affiliation{LS Theoretische Physik II, Brandenburg University of Technology, Erich-Weinert Str. 1, 03046 Cottbus, Germany} 

\author{Floren\c{t}a Costache}
\author{J\"{u}rgen Reif}
\affiliation{LS Experimentalphysik II, Brandenburg University of Technology, Universit\"{a}tsplatz 3-4, 03044 Cottbus, Germany
								 }
\date{\today}

\begin{abstract}
Costache et al. have reported recently  a new type of periodic patterns
generated at femtosecond laser ablation of transparent dielectrics (Appl. Surf.
Sci. \textbf{186}, 352(2002)). They show features known from other pattern
forming systems far from equilibrium, like point and line defects or grain
boundaries, and cannot be explained by the classical theory.  The present work
is an attempt to investigate these pattern by means of a generalized
Kuramoto-Shivashinsky equation derived from the Bradley et al. and Cuerno et
al. model for ripple formation at ion beam sputtering of surfaces.
\end{abstract}

\maketitle

%\section{INTRODUCTION}

Laser induced periodic surface structures (LIPSS) have been observed for almost
40 years \citep{lipss:univ_phenomenon} on targets made of intrinsic and
extrinsic semiconductors, metals and dielectrics, using cw to subpicosecond
laser sources with wave lengths varying from the ultraviolet up to the infrared
domain (\cite{lipss_1} and references therein,
\citep{lipss_2,lipss_3,lipss:uv}). They have therefore been considered to be a
universal phenomenon that can occur on any material that absorbs radiation,
regardless of its dielectric constant \cite{lipss:univ_phenomenon}.

The explanation accepted today on a large scale has been delivered by the group
of \citeauthor*{lipss_1} \citep{lipss_1} in \citeyear{lipss_1}, by taking into
account the details of the interaction of an electromagnetic wave with the
microscopically rough selvedge of a surface. Thus, the periodic surface
structures are the result of the interference between the incoming laser light,
the light refracted by the bulk and the light scattered by the rough selvedge
(or interface). The aspect of the surface is usually a combination of several
basic patterns. For $p$-polarized light, these are the $s$-type patterns
(Eq.~\eqref{classical:model}) perpendicular to the electric field $\vec{E}$ and
the $c$-type patterns parallel to $\vec{E}$; for $s$-polarized light, there are
$c$-type patterns perpendicular to $\vec{E}$ and other patterns with no simple
dependence on $\theta$, the angle of incidence. Excellent agreement with
experimental work that followed \citep{lipss_2,lipss_3} has been found. 

\begin{equation}
s\text{-type: }\Lambda=\frac{\lambda}{1\pm\sin\theta},\ \ \ \ \ c\text{-type: }\Lambda=\frac{\lambda}{\cos\theta}.
\label{classical:model}
\end{equation}
$\lambda$ denotes here the wave length of the laser and $\Lambda$ the
periodicity of the patterns.

Recently, \citeauthor{newstruct} have reported upon a new type of periodic
patterns generated at femtosecond laser ablation of transparent dielectrics
\citep{newstruct}.  These new structures exhibit fundamental differences when
compared to the ``classical'' periodic patterns, such as spacing almost twice
as small, independent on the wave length and on the angle of incidence of the
laser beam, but correlated with the local incident intensity. Features like
point and line defects or grain boundaries can also not be described by the
classical theory. They point rather to a self-organizing mechanism.
%The numerous bifurcations, the similarities to non-equilibrium patterns, such as sand under shallow, wavy water or the patterns on ion beam sputtered surfaces, point to a self-organizing mechanism.

The qualitative and quantitative similarities to the patterns observed in ion
beam sputtering \citep{hab:lieb}, as well as the wide acceptance of their
theoretical description as proposed by \citeauthor{bradley:harper}
\citep{bradley:harper} and \citeauthor{cuerno} \citep{cuerno}, have determined
us to use this framework for studying the new structures.

%\section{EXPERIMENTAL BACKGROUND}
The experiments \citep{newstruct,fc:ii,reif:2beams,arcdischarge} have been
carried out  under high vacuum ($<\unit{\power{10}{-7}}{\milli\bbar}$) on
freshly  cleaved single crystal slides of $\mathrm{BaF_2}$ and
$\mathrm{CaF_2}$. The laser  system generated a pulse of
\unit{120}{\femto\second} with a central wave length of \unit{800}{\nano\meter}
and intensities of $\unit{1-12\times10^{12}}{\watt\per\centi\meter\squared}$.
Additionally, the frequency of the laser radiation could be doubled and the
angle of incidence of the laser beam on the target could be varied.

\begin{figure*}
\begin{center}
\subfigure[]{\includegraphics[height=4cm]{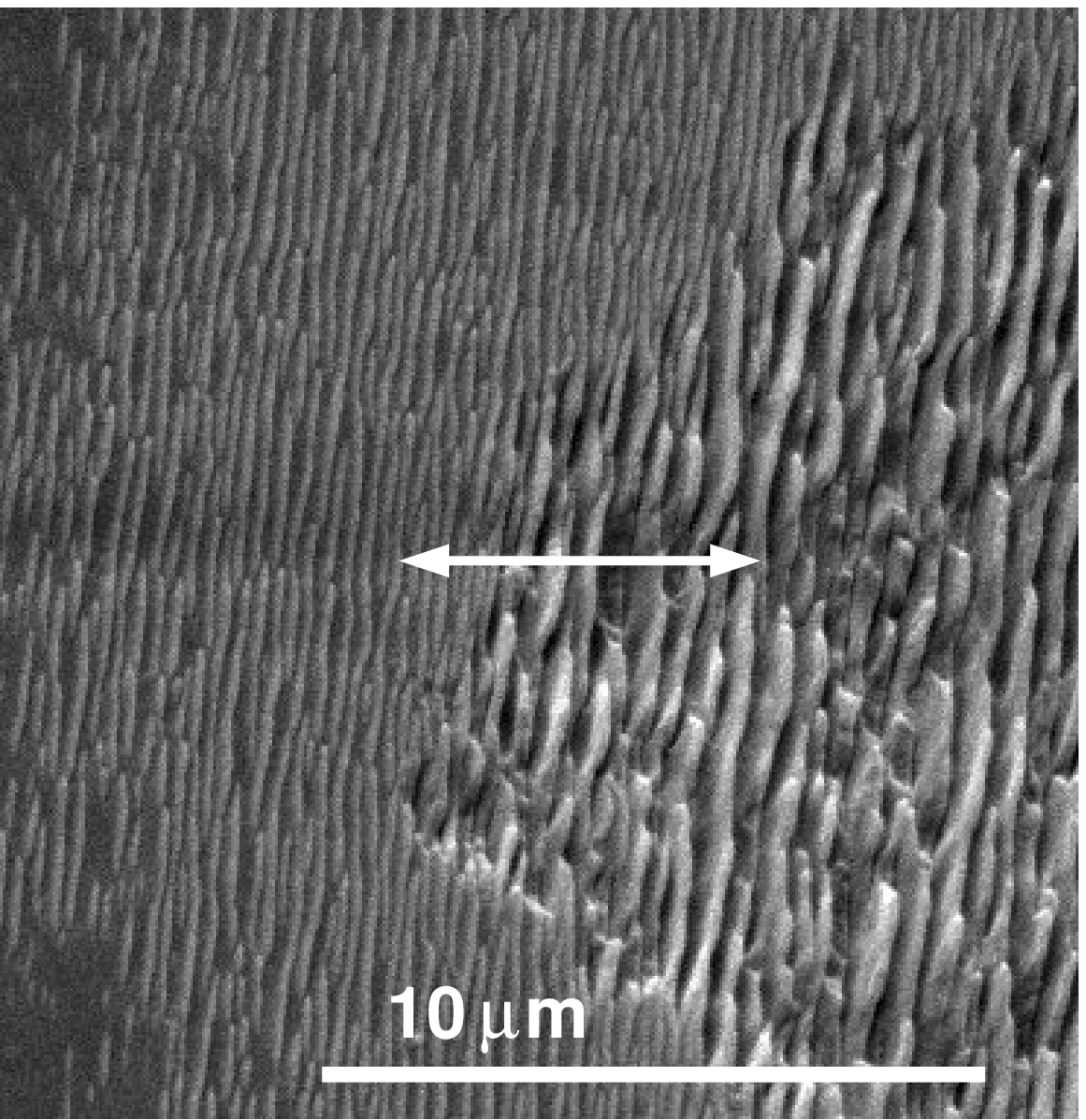}\label{laserbif_p00}}%
\hspace{0.5cm}%
\subfigure[]{\includegraphics[height=4cm]{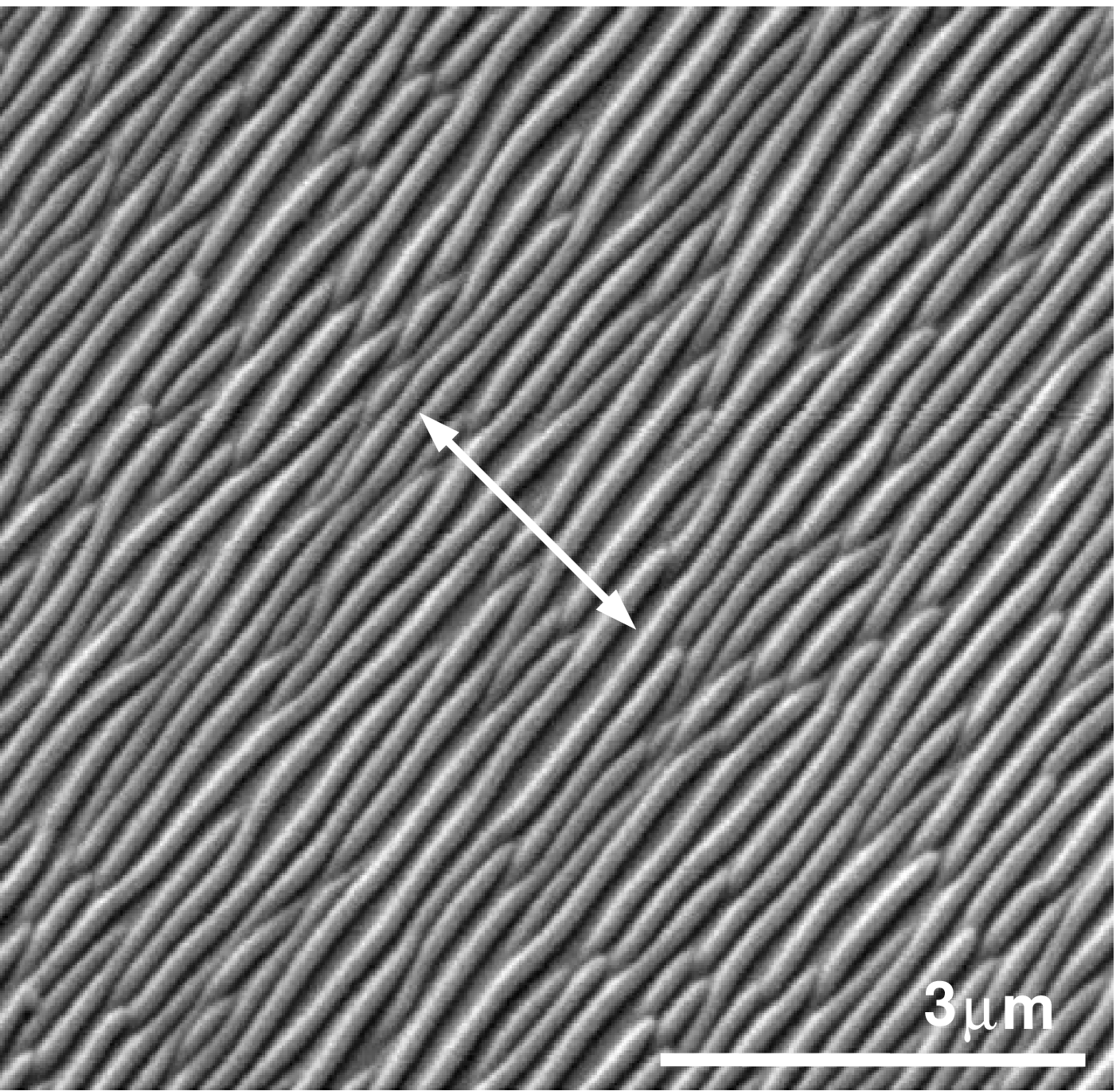}\label{laserbif_p45}}%
\hspace{0.5cm}%
\subfigure[]{\includegraphics[height=4cm]{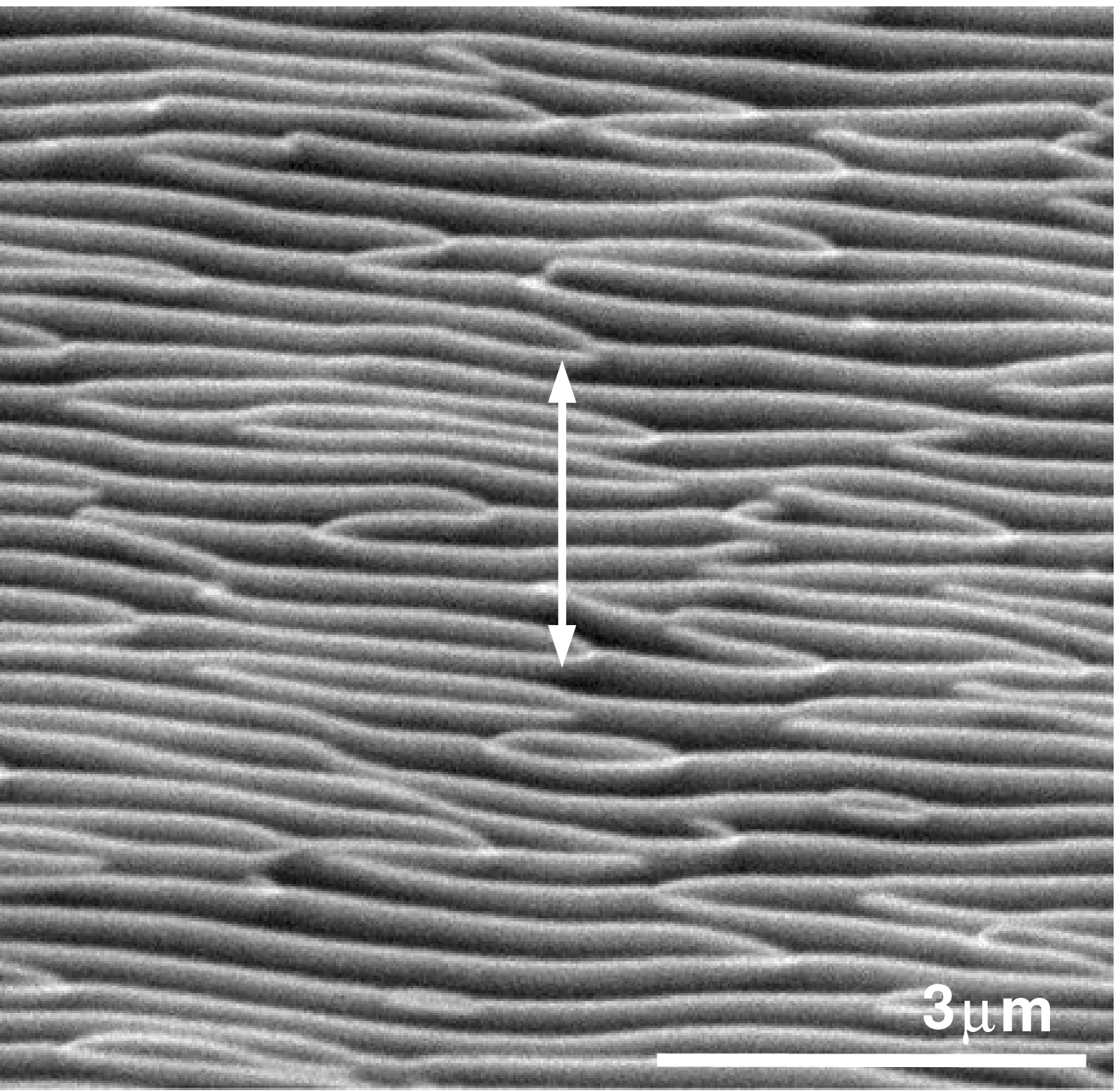}\label{laserbif_p90}}%
\end{center}
\caption{Orientation of the periodic ripple structure with respect to the laser
beam polarization. (a) Ablated spot obtained after 9200 shots at
$\unit{0.8\times\power{10}{13}}{\watt\per\centi\squaren\meter}$; (c) 5000 shots
at $\unit{1.2\times\power{10}{13}}{\watt\per\centi\squaren\meter}$;  (a), (b)
and (c) \unit{800}{\nano\meter}, \unit{45}{\degree} incidence.}%
\label{laserbif_p}
\end{figure*}

The main feature which distinguishes these new structures from the
\emph{classical} ones is the spacing. In the classical model the patterns are
uniformly distributed across the ablation area, with spacings given by
Eq.~\eqref{classical:model}. In this case, however, no dependency of the
ripples periodicity on the laser wave length or on the angle of incidence could
be observed. Moreover, the periodicity is obviously dependent on the local
intensity of the laser beam, as shown in Fig.~\ref{laserbif_p00}.  The spacing
of the ripples increases here towards the center of the ablation spot, that is,
increases with the local intensity. While Eq.~\eqref{classical:model} allows a
minimum spacing of $\approx\unit{468}{\nano\meter}$ for the situation shown in
Fig.~\ref{laserbif_p00}, the spacing at the boundary of the ablation spot is
approximatively \unit{250}{\nano\meter}.

%To further check the idea of an interference mechanism leading to surface
%ripples, \citeauthor{reif:2beams} have produced a controlled interference by
%crossing two non-collinear beams under a small angle and examined the surface
%for temporally overlapped pulses \citep{reif:2beams}.  While the Bragg's
%condition allowed for an interference pattern of spacing
%$\Lambda=\lambda/2\sin\theta\approx\unit{15}{\micro\meter}$, whose existence
%has also been checked experimentally, no evidence for it has been found in the
%surface structure.

%\section{MODEL AND SIMULATIONS}
The model presented by \citeauthor{bradley:harper} \citep{bradley:harper} and
further developed by \citeauthor{cuerno} \citep{cuerno} is based on Sigmund's
theory of sputtering \citep{sputt:i}. Here, an ion striking a solid will first
travel a certain distance $a$, called average depth of energy deposition, and
then lose its energy in a cascade of random atomic collisions. For an ion
travelling along the $z$-axis, the resulting profile of the average energy
deposition in the material ($z\leq h(x,y)$) follows in a good approximation  a
Gaussian distribution \citep{sputt:i}

\begin{equation}
F_D(\vec{r})=\frac{\epsilon}{(2\pi)^{3/2}\alpha\beta^2}
\times
\exp\left[ -\frac{[z + a]^{2}}{2\alpha^2}
    - \frac{x^2+y^2}{2\beta^2} \right].\label{gauss}
\end{equation}
$F_D(\vec{r})$ denotes here the energy deposition per unit volume, $\epsilon$
is the total energy deposited by the ion and $\alpha$ and $\beta$ are the
widths of the distribution parallel and perpendicular to the plane of
incidence, respectively. The erosion velocity at a certain point on the surface
is then proportional to the total energy deposited there by the ions within the
range $\mathcal{R}$ of the distribution \eqref{gauss}
\begin{equation}
v(O) \cong \Lambda\int_{\mathcal{R}} \Psi(\vec{r}) F_D(\vec{r}) d\vec{r},
\label{erosion:rate}
\end{equation}
where $\Psi(\vec{r})$ is the local flux of ions and $\Lambda$ is a material
constant.

Eq.~\eqref{gauss} was written in the reference frame of the incoming ion, having the origin at the point of impact $O$ and the $z$-axis pointing away from the surface, along the velocity $\vec{v}$ of the ion. In Eq.~\eqref{erosion:rate} we went over to the local reference frame, with the origin also at $O$ but with the $z$-axis pointing along the surface normal. And finally, there is the reference frame of the laboratory, our destination.

To evaluate the erosion velocity \eqref{erosion:rate}, we make
the transform $x\rightarrow a\zeta_x$, $y\rightarrow a\zeta_y$ and expand the
integrand in powers of $a/R_{X}$ and $a/R_{Y}$. $R_X$ and $R_Y$ are
the curvature radii at $O$ and are used to describe the integration region
$\mathcal{R}$
\begin{equation}
h\cong-\frac{1}{2}\left(\frac{x^2}{R_X}+\frac{y^2}{R_Y}\right)
\label{taylor:i}
\end{equation}
with $1/R_X=-\partial^2 h/\partial x^2$ and $1/R_Y=-\partial^2 h/\partial y^2$.
%\begin{equation}
%\frac{1}{R_X}=-\frac{\partial^2 h}{\partial x^2}\text{ and }
%\frac{1}{R_Y}=-\frac{\partial^2 h}{\partial y^2}
%\end{equation}
After evaluating the resulting Gaussian integrals we make the transition to the laboratory coordinate frame $(x, y, h)$ \citep{kpz, cuerno}
\begin{equation}
\frac{\partial h(x, y, t)}{\partial t}=-v(\phi, R_X, R_Y)\sqrt{1+(\nabla h)^2}
\label{lab:coord}
\end{equation}
and obtain thus for $\alpha=\beta$ \citeauthor{cuerno}'s equation of motion \citep{cuerno}
\begin{eqnarray}
\frac{\partial h}{\partial t}&=&
- v_0 + \mu\frac{\partial h}{\partial x}
+ \nu_x\frac{\partial^2 h}{\partial x^2}
+ \nu_y\frac{\partial^2 h}{\partial y^2}
+ \frac{\lambda_x}{2}\left(\frac{\partial h}{\partial x}\right)^2 +\nonumber\\
&&
\frac{\lambda_y}{2}\left(\frac{\partial h}{\partial y}\right)^2
- K\nabla^2(\nabla^2 h) + \eta(x, y, t).
\label{surfev}
\end{eqnarray}
In Eq.~\eqref{lab:coord} $\phi$ denotes the angle between the incident beam and
the surface normal at $O$. It can be expressed as a function of the angle of
incidence $\theta$ and the local gradient $\nabla h$ and can be expanded in
powers of the latter.

Eq.~\eqref{surfev} has the form of the Kuramoto-Shivashinsky (KS) equation
\citep{KS:erosion_model}.  $v_0$,  $\mu$, $\nu_x$, $\nu_y$, $\lambda_x$ and
$\lambda_y$ are functions of the angle of incidence $\theta$. The term
$-K\nabla^2(\nabla^2 h)$ accounts for the surface self-diffusion and $\eta(x,
y, t)$ is a Gaussian white noise, accounting for the stochastic arrival of the
ions on the surface. If the surface diffusion is thermally activated, the coefficient K is given by \citep{bradley:harper}
\begin{equation}
K=D_S\gamma\nu/n^2k_BT,\ D_S=D_{S_0}e^{-Q_a/k_BT}
\label{surf:diff}
\end{equation}
where $D_S$ \citep{neumann} is the surface self-diffusivity, $Q_a$ is the activation energy, $\gamma$ is the surface free energy per unit area and $\nu$ is the areal density of diffusing atoms.

In the linear regime, periodical stripes with
\begin{equation}
|\vec{k}|=\sqrt{-\nu/2K},\ \nu=\min(\nu_x, \nu_y)
\label{k}
\end{equation}
have the highest growth rate. Consequently, for $\nu=\nu_x<0$ the surface will be dominated by ripples along the $\hat{y}$-axis ($\vec{k}\parallel\hat{x}$) and for $\nu=\nu_y<0$ by ripples parallel to the $x$-axis ($\vec{k}\parallel\hat{y}$).

In the case of the laser material interaction, we assume that the photons loose
their energy in a similar stochastic process. The widths $\alpha$ and $\beta$
of the Gauss distribution \eqref{gauss} are equal because the target is
isotropic. The incident energy flux $\epsilon\Psi(\vec{r})$ is now given by
$a\gamma|\vec{S}_2\vec{n}|$, where $a$ has the meaning of an average
penetration depth, $\gamma$ is the absorption coefficient of the material and
$\vec{S}_2$ is the Poynting vector of the refracted beam. The product $a\gamma$
represents the fraction of the transmitted energy that will be absorbed, that
is, in an ionic picture, \emph{scattered} by the material. Using the Fresnel
equations, the incident flux can be written in terms of the laser intensity
$I_0=|\vec{S}_1|$ and of the transmission coefficient $T$

\begin{equation}
\epsilon\Psi(\vec{r})\rightarrow a\gamma \frac{|\vec{S}_2\vec{n}|}{|\vec{S}_1\vec{n}|}\frac{|\vec{S}_1\vec{n}|}{|\vec{S}_1|}|\vec{S}_1|=a\gamma I_0 T(\vec{r}, \phi)\cos\phi.
\label{transmission}
\end{equation}

In the next step we insert the energy flux \eqref{transmission} and the
Gaussian distribution \eqref{gauss} into Eq.~\eqref{erosion:rate} to compute
the local erosion rate. To evaluate the integral \eqref{erosion:rate}, we extend the Taylor expansion \eqref{taylor:i} to its complete form
\begin{equation}
h\cong-\frac{1}{2}\left(\frac{x^2}{R_X}+\frac{y^2}{R_Y}\right) -\frac{xy}{R_{XY}},
\label{taylor:ii}
\end{equation}
where $1/R_{XY}=-\partial^2 h/\partial x\partial y$.

We call the last term in Eq.~\eqref{taylor:ii} a ``rotational''
correction. In their original model, \citeauthor{bradley:harper}
have neglected it and could therefore only describe patterns
growing along the $x$- or the $y$-axis. With this term the stability analysis
can still be reduced to the case of \citeauthor{bradley:harper}, but in a
reference frame rotated by an angle $\Theta$ to the original one.
%\footnote{We should polish this one by regarding the laser polarization as an additional degree of freedom.}.
Following the notation style of Eq.~\eqref{surfev} and denoting by $\nu_{xy}$ the coefficient of $h_{xy}$, the angle $\Theta$ is determined by
\begin{equation}
\nu_{xy}\cos2\Theta = (\nu_x - \nu_y)\sin2\Theta.
\label{Theta}
\end{equation}

We perform the transition \eqref{lab:coord} to the coordinate frame of the laboratory and scale the time and the space according to
\begin{equation}
x=ax',\ y=ay',\ t= \frac{\sqrt{2\pi}a}{\Lambda\gamma I_0}.
\label{scaling}
\end{equation}
to obtain
\begin{eqnarray}
\dot{h}&=&-v_0+\mu h_x + \nu_x h_{xx} + \nu_y h_{yy} + \nu_{xy} h_{xy} + \frac{1}{2}\lambda_x h_x^2 \nonumber \\
&& + \frac{1}{2}\lambda_y h_y^2 + \nu_{10100} h_x h_{xx} + \nu_{10010} h_x h_{yy} + \nonumber \\
&& \nu_{10001} h_x h_{xy} + \nu_{00200}h_{xx}^2 + \nu_{00020}h_{yy}^2 + \nu_{00002} h_{xy}^2 \nonumber \\
&& + \nu_{00110}h_{xx}h_{yy} + \nu_{00101}h_{xx}h_{xy} + \nu_{00011}h_{yy}h_{xy} \nonumber \\
&&-B\Delta^2h.
\label{ks:laser}
\end{eqnarray}
%All the coefficients of Eq.~\eqref{ks:laser} except $B$  have the common factor $F=-\sigma\exp\left(-1/2\sigma^2(n^2-\sin^2\theta)/n^2\right)$. In particular, we write $\nu_{\ldots}$ as $\nu_{\ldots}=-F\Gamma_{\ldots}(\sigma, \varphi, \theta, n)$.

The tuples $ijklm$ in $\nu_{ijklm}$ describe
the product $h_x^i h_y^j h_{xx}^k h_{yy}^l h_{xy}^m$ these coefficients belong
to. Except for $B$, all the coefficients in Eq.~\eqref{ks:laser} have the form $\{,\}=\sigma\exp[-\frac{\sigma^2}{2} \cos^2\theta] f_{\{,\}}(n, \theta, \varphi)$, with $\sigma=a/\alpha$.

%\section{RESULTS}

\begin{figure}
\begin{center}
\includegraphics[width=7cm]{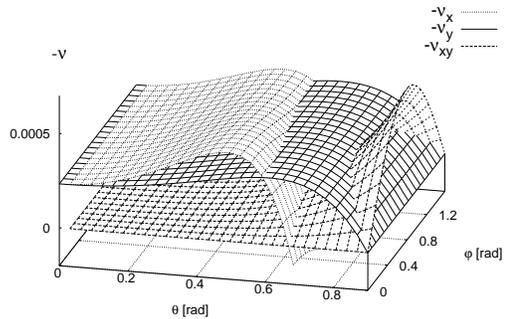}%
\end{center}
\caption{Plot of $\nu_x$, $\nu_y$ and $\nu_{xy}$ against the angle of incidence $\theta$ and the polarization $\varphi$. $\sigma = $ 4, $n$ = 1.47}\label{gammasel}
\end{figure}

Fig.~\ref{gammasel} shows a 3D plot of $-\nu_x$, $-\nu_y$ and $-\nu_{xy}$
against the angle of incidence $\theta$ and the polarization $\varphi$. We
chose $\sigma$=4, $B=1.2\cdot10^{-4}$ and $n$=1.47, the refraction index of
$\mathrm{BaF_2}$ at $\lambda\sim\unit{750}{\nano\meter}$. As shown in the
previous section, the larger of $-\nu_x>0$ and $-\nu_y>0$ determines the
prevailing pattern and $\nu_{xy}$ will rotate this pattern by an angle $\Theta$
(Eq.~\eqref{Theta}).

We see that there is actually only a small region between $\theta_{\text{min}}$ and $\theta_{\text{max}}$, where the ripple orientation changes with the light polarization. In our case $\theta_{\text{min}}$ was \unit{29.64}{\degree} and $\theta_{\text{max}}$ = \unit{34.04}{\degree}. For angles smaller than $\theta_{\text{min}}$ the ripples should always be perpendicular to the surface component of the laser beam, where as for angles larger than $\theta_{\text{max}}$ they should be parallel.  The width of this region
%$\theta_{\text{max}}-\theta_{\text{min}}$
is usually between 1 and 5 degrees, increasing with increasing $n$. Its position varies slightly with $\sigma$, but strongly with $n$. Smaller $n$'s and larger $\sigma$'s will move it towards angles of up to \unit{50-60}{\degree}.

%Lower $n$ are motivated by ...

\begin{figure}
\begin{center}
\includegraphics[width=4cm]{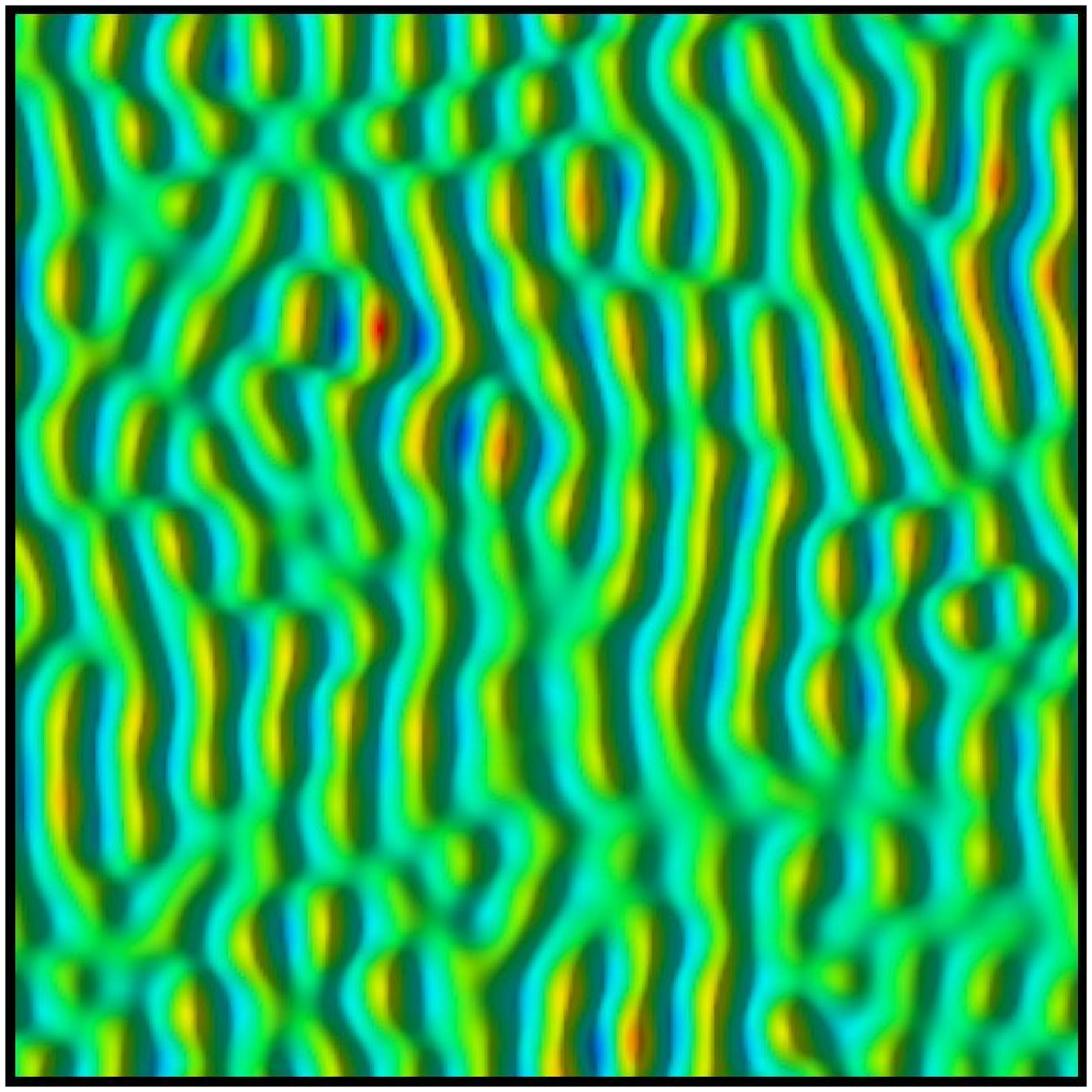}%
\hspace{0.3cm}%
\includegraphics[width=4cm]{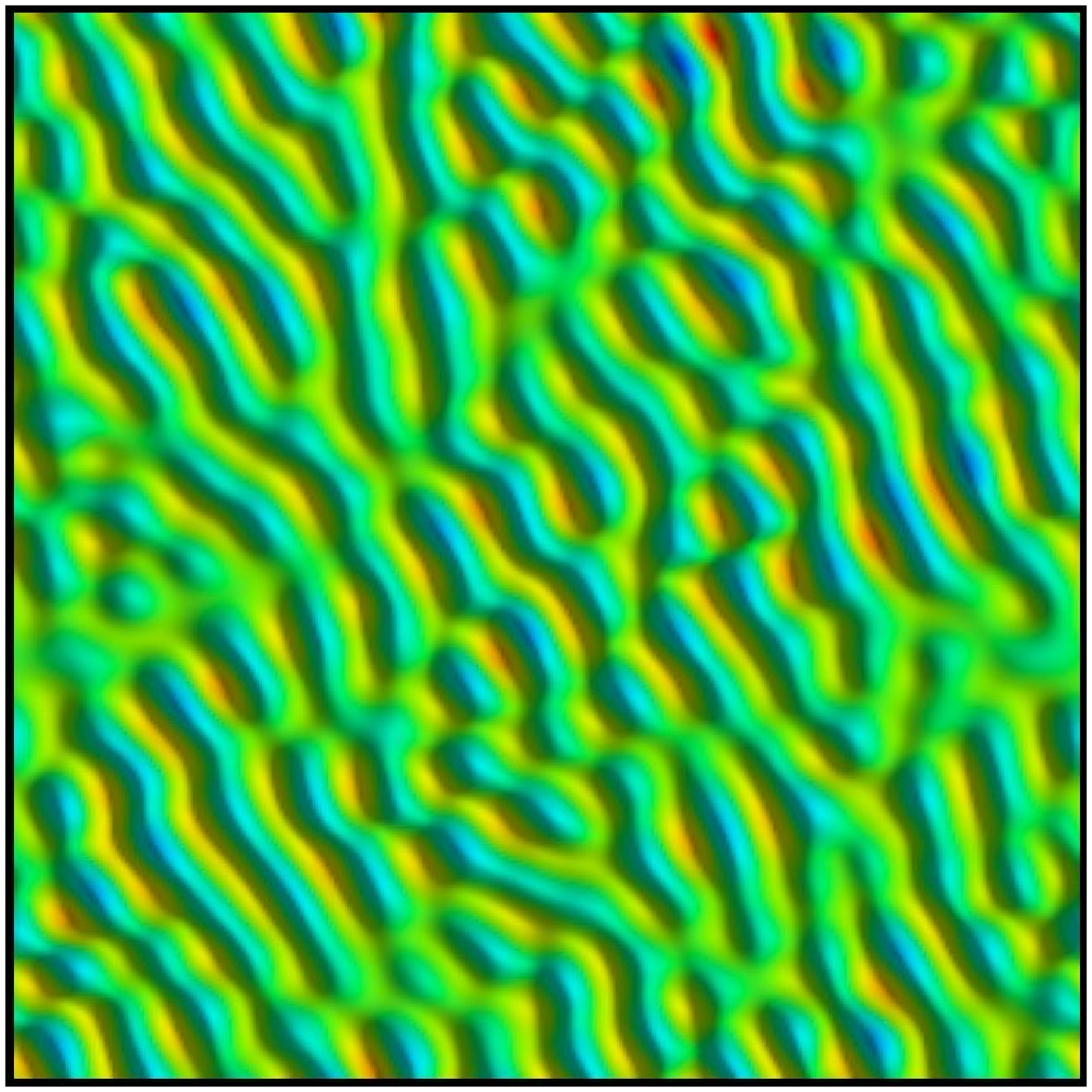}\\
\includegraphics[width=4cm]{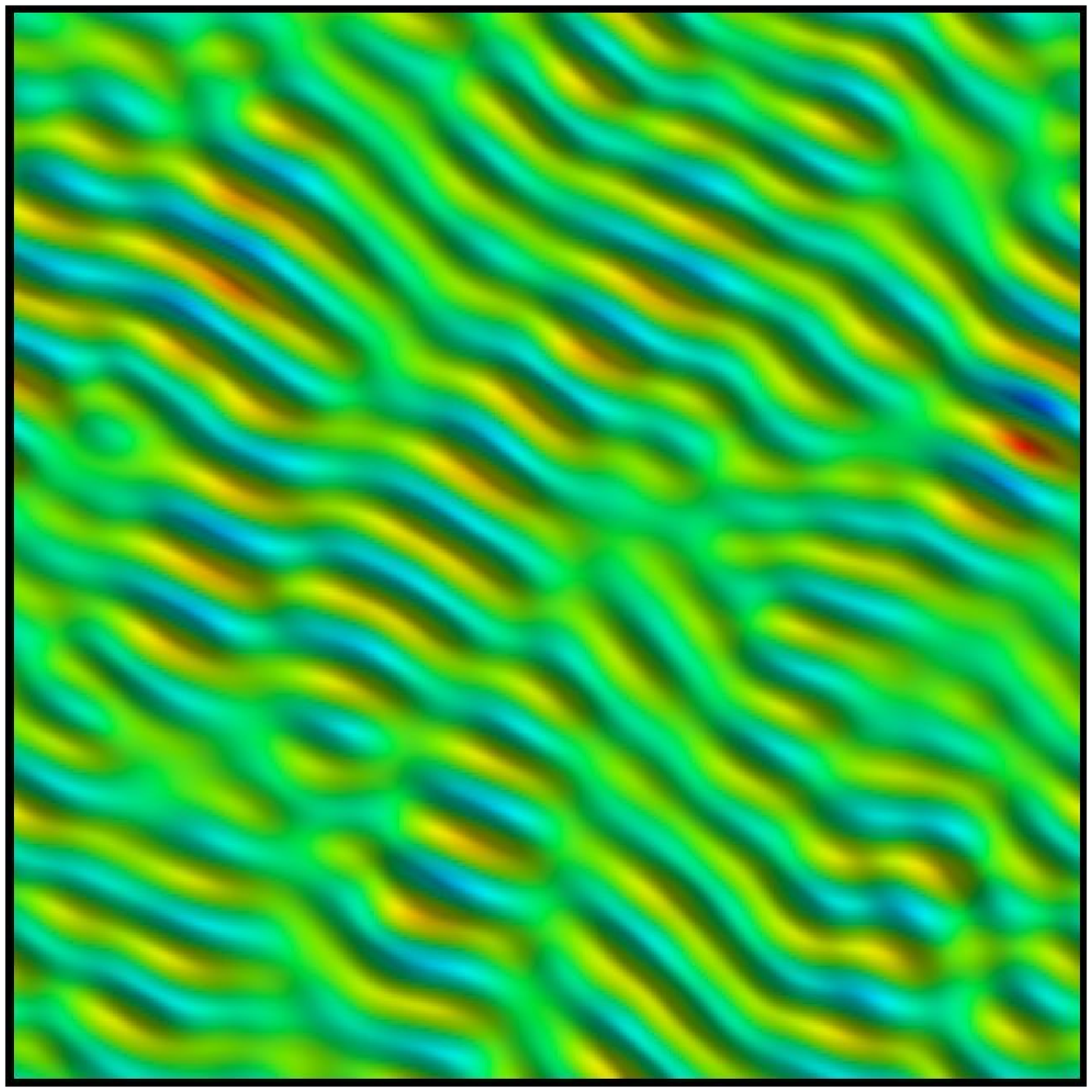}%
\hspace{0.3cm}%
\includegraphics[width=4cm]{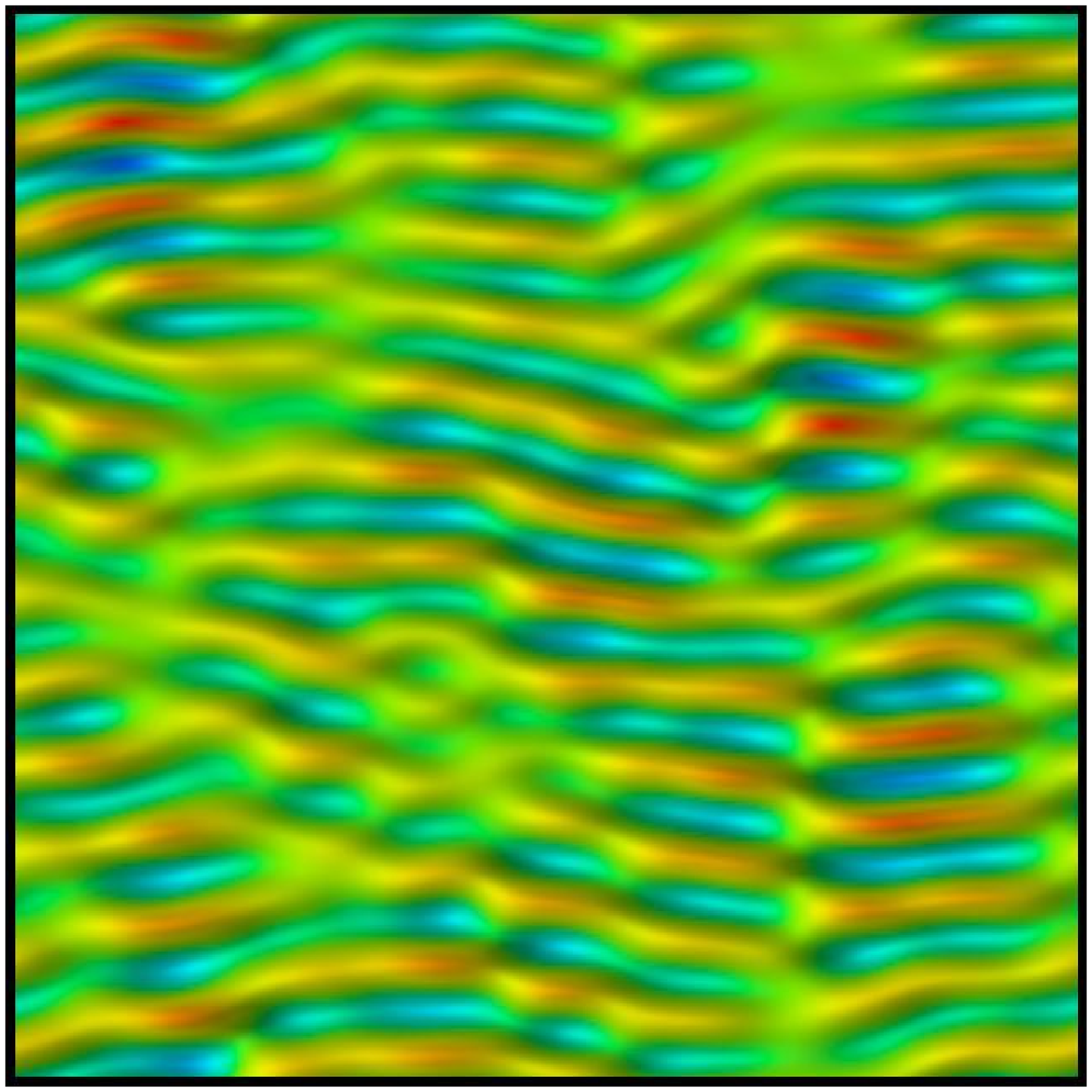}
\end{center}
\caption{Surface topography for $\theta$=\unit{31.8}{\degree} and $\varphi$=\unit{0}{\degree}, \unit{30}{\degree}, \unit{60}{\degree} and \unit{90}{\degree} respectively (upper-left to lower-right).}\label{surf:topography}
\end{figure}

%\section{DISCUSSION}

The present work is an attempt to investigate the occurrence of newly
discovered laser induced periodic surface structures \citep{newstruct} by means
of the Bradley et~al. \citep{bradley:harper} and Cuerno et~al. \citep{cuerno}
model. Based on Sigmund's stochastical theory of sputtering \citep{sputt:i},
this model has managed to explain in an unified framework most of the dynamic
and scaling behaviors observed experimentally at ion bombardment of surfaces.
It eventually traces the origin of ripple formation down to the instability
caused by the competition between surface roughening (so called negative
surface tension) and surface diffusion (the positive surface tension).

The use of the Fresnel equations and the inclusion of the $\nu_{xy}(\partial^2
h/\partial x \partial y)$ term to the Kuramoto-Sivashinsky equation, have
allowed us to confirm the experimentally observed countinous change of the
ripple orientation with the laser polarization. However, our model predicts
that this should only happen for angles of incidence confined in a small range
between $\theta_{\text{min}}$ and $\theta_{\text{max}}$. In our simulations
this range corresponded to \unit{29\,-\,34}{\degree}, but could move up to \unit{53\,-\,55}{\degree}.
%Experiments at other angles of incidence would be therefore of great help,
%since \citeauthor{newstruct} have performed their experiments only at
%\unit{0}{\degree} and \unit{45}{\degree} incidence \citep{newstruct, fc:ii}.

The role of the local laser intensity can also be explained. According to
Eq.~\eqref{surf:diff}, for local temperatures $T_{\text{local}}$ below $Q_a$,
$K$ is a monotonically increasing function of $T_{\text{local}}$ and thus of
the laser intensity $I_{\text{local}}$.  A first estimation yields $Q_a >
\unit{0.53}{\electronvolt}$
\footnote{$Q_a$ is additively composed of $H_{Fa}$ and $H_{Ma}$, the formation
and migration energy of an adatom \citep{neumann}.  We have numerically evaluated $H_{Ma}$
for the (110) surface of $\mathrm{BaF_2}$ and
obtained \unit{0.53}{\electronvolt} \citep{feynman}.}
, much larger than the melting point of $\mathrm{BaF}_2$
($\unit{1280}{\celsius}\,\approx\,\unit{0.13}{\electronvolt}$). The periodicity
$2\pi/|\vec{k}|$ of the ripples (Eq.~\eqref{k}) is thus an increasing function
of $I_{\text{local}}$, just like in Fig.~\ref{laserbif_p00}.

At normal incidence, all patterns with the wave vector satisfying
$|\vec{k}|=\sqrt{-\nu/2B}$ can occur with equal probability. The symmetry
break has therefore been looked for and traces of it have been found in the
3$^{\mathrm{rd}}$ order terms. However, further investigation is needed, as we have not
been able to simulate any ripples yet.

%The direct comparison of the simulated wave lengths and times with the
%experiment is not yet possible. Not only is there a great number of unknown
%parameters, $a$, $\alpha$ (or $\sigma=a/\alpha$), $\Lambda$ and $K$, but the lack of a comprehensive study of their time behaviour makes it impossible.

It is well known that the spacing of the structures described by the KS
equation increases with time according to $\lambda\sim t^{\gamma}$. The value
of $\gamma$ varies with different experimental conditions
\citep{ionhexagons,hab:lieb,hab:bol} and different numerical simulations
\citep{kpz, KS:erosion_model, KS:numerical_analysis:2D}. Considering that the
exposure times varied across the experiments, an unknown value of $\gamma$
makes the direct quantitative comparison of the numerical simulations with the
experiment impossible. A comprehensive study of the time behavior of the
structures is therefore a further requirement for their understanding.

\citeauthor{ionhexagons} have recently shown that the compound nature of a
material can considerably influence the scaling behavior of the surface
structures in the early stages of ion sputtering \citep{ionhexagons}.
Considering the extreme conditions of the laser ablation, the loss of atoms is
with high probability different for Ba and F, so that a spacially varying
concentration of adatoms on the surface would arise.
\citeauthor{adatoms_gradient} have shown that surface diffusion driven by a
concentration gradient of the adatoms is able to generate structure coarsening
\citep{adatoms_gradient}. According to \citeauthor{ionhexagons}, this type of
coarsening would be one of the causes for the deviation from the KS predicted
scaling behavior they have observed in the early stages of $\mathrm{Ar}^+$
sputtering of InP surfaces.

Further experiments with rotating targets or circular polarized light could
also confirm the common nature of this new type of LIPSS and of the ion
sputtering processes. Studying the topography of simultaneously rotated and
$\mathrm{Ar}^+$ ion sputtered InP surfaces, \citeauthor{ionhexagons} have been
able to find structures of highly hexagonal symmetry \citep{ionhexagons}.

%\begin{acknowledgments}
I. Georgescu would like to thank Mario de Menech for helpful discussions.
%\end{acknowledgments}

%\nocite{*}
\bibliographystyle{apsrev}
\bibliography{literature}

\begin{thebibliography}{21}
\expandafter\ifx\csname natexlab\endcsname\relax\def\natexlab#1{#1}\fi
\expandafter\ifx\csname bibnamefont\endcsname\relax
  \def\bibnamefont#1{#1}\fi
\expandafter\ifx\csname bibfnamefont\endcsname\relax
  \def\bibfnamefont#1{#1}\fi
\expandafter\ifx\csname citenamefont\endcsname\relax
  \def\citenamefont#1{#1}\fi
\expandafter\ifx\csname url\endcsname\relax
  \def\url#1{\texttt{#1}}\fi
\expandafter\ifx\csname urlprefix\endcsname\relax\def\urlprefix{URL }\fi
\providecommand{\bibinfo}[2]{#2}
\providecommand{\eprint}[2][]{\url{#2}}

\bibitem[{\citenamefont{van Driel et~al.}(1982)\citenamefont{van Driel, Sipe,
  and Young}}]{lipss:univ_phenomenon}
\bibinfo{author}{\bibfnamefont{H.~M.} \bibnamefont{van Driel}},
  \bibinfo{author}{\bibfnamefont{J.~E.} \bibnamefont{Sipe}}, \bibnamefont{and}
  \bibinfo{author}{\bibfnamefont{J.~F.} \bibnamefont{Young}},
  \bibinfo{journal}{Phys. Rev. Lett.} \textbf{\bibinfo{volume}{49}},
  \bibinfo{pages}{1955} (\bibinfo{year}{1982}).

\bibitem[{\citenamefont{Sipe et~al.}(1983)\citenamefont{Sipe, Young, Preston,
  and van Driel}}]{lipss_1}
\bibinfo{author}{\bibfnamefont{J.~E.} \bibnamefont{Sipe}},
  \bibinfo{author}{\bibfnamefont{J.~F.} \bibnamefont{Young}},
  \bibinfo{author}{\bibfnamefont{J.~S.} \bibnamefont{Preston}},
  \bibnamefont{and} \bibinfo{author}{\bibfnamefont{H.~M.} \bibnamefont{van
  Driel}}, \bibinfo{journal}{Phys. Rev. B} \textbf{\bibinfo{volume}{27}},
  \bibinfo{pages}{1141} (\bibinfo{year}{1983}).

\bibitem[{\citenamefont{Young et~al.}(1983)\citenamefont{Young, Preston, van
  Driel, and Sipe}}]{lipss_2}
\bibinfo{author}{\bibfnamefont{J.~F.} \bibnamefont{Young}},
  \bibinfo{author}{\bibfnamefont{J.~S.} \bibnamefont{Preston}},
  \bibinfo{author}{\bibfnamefont{H.~M.} \bibnamefont{van Driel}},
  \bibnamefont{and} \bibinfo{author}{\bibfnamefont{J.~E.} \bibnamefont{Sipe}},
  \bibinfo{journal}{Phys. Rev. B} \textbf{\bibinfo{volume}{27}},
  \bibinfo{pages}{1155} (\bibinfo{year}{1983}).

\bibitem[{\citenamefont{Young et~al.}(1984)\citenamefont{Young, Sipe, and van
  Driel}}]{lipss_3}
\bibinfo{author}{\bibfnamefont{J.~F.} \bibnamefont{Young}},
  \bibinfo{author}{\bibfnamefont{J.~E.} \bibnamefont{Sipe}}, \bibnamefont{and}
  \bibinfo{author}{\bibfnamefont{H.~M.} \bibnamefont{van Driel}},
  \bibinfo{journal}{Phys. Rev. B} \textbf{\bibinfo{volume}{30}},
  \bibinfo{pages}{2001} (\bibinfo{year}{1984}).

\bibitem[{\citenamefont{Clark and Emmony}(1989)}]{lipss:uv}
\bibinfo{author}{\bibfnamefont{S.~E.} \bibnamefont{Clark}} \bibnamefont{and}
  \bibinfo{author}{\bibfnamefont{D.~C.} \bibnamefont{Emmony}},
  \bibinfo{journal}{Phys. Rev. B} \textbf{\bibinfo{volume}{40}},
  \bibinfo{pages}{2031} (\bibinfo{year}{1989}).

\bibitem[{\citenamefont{Costache et~al.}(2002)\citenamefont{Costache, Henyk,
  and Reif}}]{newstruct}
\bibinfo{author}{\bibfnamefont{F.}~\bibnamefont{Costache}},
  \bibinfo{author}{\bibfnamefont{M.}~\bibnamefont{Henyk}}, \bibnamefont{and}
  \bibinfo{author}{\bibfnamefont{J.}~\bibnamefont{Reif}},
  \bibinfo{journal}{Appl. Surf. Sci.} \textbf{\bibinfo{volume}{186}},
  \bibinfo{pages}{352} (\bibinfo{year}{2002}).

\bibitem[{\citenamefont{Bradley and Harper}(1988)}]{bradley:harper}
\bibinfo{author}{\bibfnamefont{R.~M.} \bibnamefont{Bradley}} \bibnamefont{and}
  \bibinfo{author}{\bibfnamefont{J.~M.~E.} \bibnamefont{Harper}},
  \bibinfo{journal}{J. Vac. Sci. Technol. A} \textbf{\bibinfo{volume}{6}},
  \bibinfo{pages}{2390} (\bibinfo{year}{1988}).

\bibitem[{\citenamefont{Cuerno and Barab\'{a}si}(1995)}]{cuerno}
\bibinfo{author}{\bibfnamefont{R.}~\bibnamefont{Cuerno}} \bibnamefont{and}
  \bibinfo{author}{\bibfnamefont{A.-L.} \bibnamefont{Barab\'{a}si}},
  \bibinfo{journal}{Phys. Rev. Lett.} \textbf{\bibinfo{volume}{74}},
  \bibinfo{pages}{4746} (\bibinfo{year}{1995}).

\bibitem[{\citenamefont{Costache et~al.}(2003)\citenamefont{Costache, Henyk,
  and Reif}}]{fc:ii}
\bibinfo{author}{\bibfnamefont{F.}~\bibnamefont{Costache}},
  \bibinfo{author}{\bibfnamefont{M.}~\bibnamefont{Henyk}}, \bibnamefont{and}
  \bibinfo{author}{\bibfnamefont{J.}~\bibnamefont{Reif}},
  \bibinfo{journal}{Appl. Surf. Sci.} \textbf{\bibinfo{volume}{203}},
  \bibinfo{pages}{486} (\bibinfo{year}{2003}).

\bibitem[{\citenamefont{Reif et~al.}(2002)\citenamefont{Reif, Costache, Henyk,
  and Pandelov}}]{reif:2beams}
\bibinfo{author}{\bibfnamefont{J.}~\bibnamefont{Reif}},
  \bibinfo{author}{\bibfnamefont{F.}~\bibnamefont{Costache}},
  \bibinfo{author}{\bibfnamefont{M.}~\bibnamefont{Henyk}}, \bibnamefont{and}
  \bibinfo{author}{\bibfnamefont{S.~V.} \bibnamefont{Pandelov}},
  \bibinfo{journal}{Appl. Surf. Sci.} \textbf{\bibinfo{volume}{197-198}},
  \bibinfo{pages}{891} (\bibinfo{year}{2002}).

\bibitem[{\citenamefont{Henyk et~al.}(1999)\citenamefont{Henyk, Vogel,
  Wolfframm, Tempel, and Reif}}]{arcdischarge}
\bibinfo{author}{\bibfnamefont{M.}~\bibnamefont{Henyk}},
  \bibinfo{author}{\bibfnamefont{N.}~\bibnamefont{Vogel}},
  \bibinfo{author}{\bibfnamefont{D.}~\bibnamefont{Wolfframm}},
  \bibinfo{author}{\bibfnamefont{A.}~\bibnamefont{Tempel}}, \bibnamefont{and}
  \bibinfo{author}{\bibfnamefont{J.}~\bibnamefont{Reif}},
  \bibinfo{journal}{Appl. Phys. A} \textbf{\bibinfo{volume}{69}},
  \bibinfo{pages}{355} (\bibinfo{year}{1999}).

\bibitem[{\citenamefont{Sigmund}(1969)}]{sputt:i}
\bibinfo{author}{\bibfnamefont{P.}~\bibnamefont{Sigmund}},
  \bibinfo{journal}{Physical Review} \textbf{\bibinfo{volume}{194}},
  \bibinfo{pages}{383} (\bibinfo{year}{1969}).

\bibitem[{\citenamefont{Kardar et~al.}(1986)\citenamefont{Kardar, Parisi, and
  Zhang}}]{kpz}
\bibinfo{author}{\bibfnamefont{M.}~\bibnamefont{Kardar}},
  \bibinfo{author}{\bibfnamefont{G.}~\bibnamefont{Parisi}}, \bibnamefont{and}
  \bibinfo{author}{\bibfnamefont{Y.-C.} \bibnamefont{Zhang}},
  \bibinfo{journal}{Phys. Rev. Lett.} \textbf{\bibinfo{volume}{56}},
  \bibinfo{pages}{889} (\bibinfo{year}{1986}).

\bibitem[{\citenamefont{Lauritsen et~al.}(1996)\citenamefont{Lauritsen, Cuerno,
  and Makse}}]{KS:erosion_model}
\bibinfo{author}{\bibfnamefont{K.~B.} \bibnamefont{Lauritsen}},
  \bibinfo{author}{\bibfnamefont{R.}~\bibnamefont{Cuerno}}, \bibnamefont{and}
  \bibinfo{author}{\bibfnamefont{H.~A.} \bibnamefont{Makse}},
  \bibinfo{journal}{Phys. Rev. E} \textbf{\bibinfo{volume}{54}},
  \bibinfo{pages}{3577} (\bibinfo{year}{1996}).

\bibitem[{\citenamefont{Neumann and Hirschwald}(1972)}]{neumann}
\bibinfo{author}{\bibfnamefont{G.}~\bibnamefont{Neumann}} \bibnamefont{and}
  \bibinfo{author}{\bibfnamefont{W.}~\bibnamefont{Hirschwald}},
  \bibinfo{journal}{Zeitschrift f{\"{u}}r Physikalische Chemie Neue Folge}
  \textbf{\bibinfo{volume}{81}}, \bibinfo{pages}{163} (\bibinfo{year}{1972}).

\bibitem[{\citenamefont{Frost et~al.}(2000)\citenamefont{Frost, Schindler, and
  Bigl}}]{ionhexagons}
\bibinfo{author}{\bibfnamefont{F.}~\bibnamefont{Frost}},
  \bibinfo{author}{\bibfnamefont{A.}~\bibnamefont{Schindler}},
  \bibnamefont{and} \bibinfo{author}{\bibfnamefont{F.}~\bibnamefont{Bigl}},
  \bibinfo{journal}{Phys. Rev. Lett.} \textbf{\bibinfo{volume}{85}},
  \bibinfo{pages}{4116} (\bibinfo{year}{2000}).

\bibitem[{\citenamefont{Habenicht and Lieb}(2002)}]{hab:lieb}
\bibinfo{author}{\bibfnamefont{S.}~\bibnamefont{Habenicht}} \bibnamefont{and}
  \bibinfo{author}{\bibfnamefont{K.~P.} \bibnamefont{Lieb}},
  \bibinfo{journal}{Phys. Rev. B} \textbf{\bibinfo{volume}{65}},
  \bibinfo{pages}{115327} (\bibinfo{year}{2002}).

\bibitem[{\citenamefont{Habenicht et~al.}(1999)\citenamefont{Habenicht, Bolse,
  and Lieb}}]{hab:bol}
\bibinfo{author}{\bibfnamefont{S.}~\bibnamefont{Habenicht}},
  \bibinfo{author}{\bibfnamefont{W.}~\bibnamefont{Bolse}}, \bibnamefont{and}
  \bibinfo{author}{\bibfnamefont{K.~P.} \bibnamefont{Lieb}},
  \bibinfo{journal}{Phys. Rev. B} \textbf{\bibinfo{volume}{60}},
  \bibinfo{pages}{R2200} (\bibinfo{year}{1999}).

\bibitem[{\citenamefont{Drotar et~al.}(1999)\citenamefont{Drotar, Zhao, Lu, and
  Wang}}]{KS:numerical_analysis:2D}
\bibinfo{author}{\bibfnamefont{J.~T.} \bibnamefont{Drotar}},
  \bibinfo{author}{\bibfnamefont{Y.-P.} \bibnamefont{Zhao}},
  \bibinfo{author}{\bibfnamefont{T.-M.} \bibnamefont{Lu}}, \bibnamefont{and}
  \bibinfo{author}{\bibfnamefont{G.-C.} \bibnamefont{Wang}},
  \bibinfo{journal}{Phys. Rev. E} \textbf{\bibinfo{volume}{59}},
  \bibinfo{pages}{177} (\bibinfo{year}{1999}).

\bibitem[{\citenamefont{Mayr et~al.}(1999)\citenamefont{Mayr, Moske, and
  Samwer}}]{adatoms_gradient}
\bibinfo{author}{\bibfnamefont{S.~G.} \bibnamefont{Mayr}},
  \bibinfo{author}{\bibfnamefont{M.}~\bibnamefont{Moske}}, \bibnamefont{and}
  \bibinfo{author}{\bibfnamefont{K.}~\bibnamefont{Samwer}},
  \bibinfo{journal}{Phys. Rev. B} \textbf{\bibinfo{volume}{60}},
  \bibinfo{pages}{16950} (\bibinfo{year}{1999}).

\bibitem[{\citenamefont{Feynman et~al.}(1964)\citenamefont{Feynman, Leighton,
  and Sands}}]{feynman}
\bibinfo{author}{\bibfnamefont{R.}~\bibnamefont{Feynman}},
  \bibinfo{author}{\bibfnamefont{R.~B.} \bibnamefont{Leighton}},
  \bibnamefont{and} \bibinfo{author}{\bibfnamefont{M.~L.} \bibnamefont{Sands}},
  \emph{\bibinfo{title}{The Feynman lectures on physics}},
  vol.~\bibinfo{volume}{II} (\bibinfo{publisher}{Addison-Wesley},
  \bibinfo{year}{1964}), \bibinfo{note}{8-3 The electrostatic energy of an
  ionic crystal}.

\end{thebibliography}

\end{document}